\documentstyle[12pt,epsf,bbm]{article} 
\newcommand{\be}{\begin{equation}} 
\newcommand{\en}{\end{equation}}
\newcommand{\bea}{\begin{eqnarray}}
\newcommand{\ena}{\end{eqnarray}}
\newcommand{\Det}{\hbox{Det}}
\newcommand{\hbo}{\hbox to 1 true cm {\hfill } } 
\newcommand{\tr}{\hbox{tr} \, }

\def\dslash{\partial\kern-.6em\slash}
\def\kslash{k\kern-.5em\slash}
\def\pslash{p\kern-.4em\slash}
\def\Dslash{D\kern-.6em\slash}
\def\Aslash{A\kern-.6em\slash}
\def\Vslash{V\kern-.7em\slash}
\def\vslash{v\kern-.5em\slash}
\def\rslash{r\kern-.5em\slash}
\def\qslash{q\kern-.5em\slash}

\begin{document} 
\vglue 1truecm
  
\vbox{ KIAS-P99025 
\hfill July 15, 1999 
}
  
\vfil
\centerline{\large\bf Lattice Yang-Mills theory at finite densities } 
\centerline{\large\bf of heavy quarks $^*$ } 
  
\bigskip
\centerline{ Kurt Langfeld$^{\, a,b}$ and Gwansoo Shin$^{\, a,c}$ } 
\vspace{1 true cm} 
\centerline{ $^a$  School of Physics, Korea Institute for Advanced
Study}
\centerline{ Seoul 130-012, Korea }
\bigskip
\centerline{ $^b$ Institut f\"ur Theoretische Physik, Universit\"at 
   T\"ubingen }
\centerline{D--72076 T\"ubingen, Germany}
\bigskip 
\centerline{ $^c$ 
Department of Physics and Center for Theoretical Physics, } 
\centerline{ Seoul National University, Seoul 151-742, Korea }

\vfil
\begin{abstract}

SU($N_c$) Yang-Mills theory is investigated at finite densities of $N_f$ 
heavy quark flavors. The calculation of the (continuum) quark determinant 
in the large-mass limit is performed by analytic methods and results in 
an effective gluonic action. This action is then subject to a lattice 
representation of the gluon fields and computer simulations. 
The approach maintains the same number of quark degrees of freedom 
as in the continuum formulation and a physical heavy quark limit (to be 
contrasted with the quenched approximation $N_f \rightarrow 0$). The proper 
scaling towards the continuum limit is manifest. We study the partition 
function for given values of the chemical potential as well as the partition 
function which is projected onto a definite baryon number. First numerical 
results for an SU(2) gauge theory are presented. We briefly discuss the 
breaking of the color-electric string at finite densities and 
shed light onto the origin of the overlap problem inherent in the Glasgow 
approach.

\end{abstract}

\vfil
\hrule width 5truecm
\vskip .2truecm
\begin{quote} 
\end{quote}
\eject

\section{Introduction} 

The next generation of particle accelerators (RHIC, LHC), which will 
start operating at the beginning of the next millennium, will 
probe the deconfined regime of QCD, the theory of strong interactions, and 
might reveal  exotic states of hadronic matter which appear under extreme 
conditions, i.e.,  temperature and density. Due to a significant increase 
in computational power in the recent past, numerical simulations of 
lattice QCD have provided insights into the high temperature and 
zero density phase and have predicted a series of interesting 
phenomena~\cite{kan98}, such as deconfinement and restoration of chiral 
symmetry. Unfortunately, an adequate description of finite 
density hadron matter is still lacking due to conceptual problems 
in setting up an appropriate ''statistical'' measure which can be 
handled in computer simulations. 

\vskip 0.3cm 
The generic approach to Yang-Mills thermodynamics at finite densities is based 
on the introduction of a non-zero chemical potential. In the case of a 
SU(2) gauge group, the fermion determinant is real and can be included 
in the probabilistic measure. Numerical simulations can be performed by 
using standard algorithms~\cite{dua87}, although this numerical 
approach consumes a lot of computer time due to the non-local nature 
of the action. Recent progress for the case of a SU(2) gauge group 
can be found in~\cite{sim99}. In the case of a SU(3) gauge group, 
the fermion determinant acquires imaginary parts for a non-vanishing 
chemical potential and cannot be considered to be part of the probabilistic 
measure. The most prominent example to circumvent this conceptual 
difficulty is the so-called Glasgow algorithm~\cite{bar92}. There, 
the fermion determinant is considered to be part of the correlation 
function to be calculated, and the probabilistic measure of 
zero-density Yang-Mills theory is used to generate the gauge field 
configurations. However, it turns out that this approach suffers from 
the so-called ''overlap'' problem implying that for realistic lattice 
sizes an unrealistic number of Monte-Carlo steps is necessary to 
achieve reliable results~\cite{bar99}. 

\vskip 0.3cm 
In order to alleviate this problem, the so-called quenched approximation, 
i.e.,  the limit $N_f \rightarrow 0$, where $N_f$ is the number 
of quark flavors, greatly reduces the numerical task for practical 
calculations. While in the case of real QCD one expects a drastic change 
in the hadron density for a chemical potential $ \mu = \mu _{onset} 
\approx m_B/3$ ($m_B$ is the baryon mass), the onset value $\mu _{onset} $ 
which is extracted from quenched lattice QCD seems to be unnaturally 
small~\cite{bar86}. Subsequently, it turned out that 
the quenched approximation, i.e.,  the limit $N_f \rightarrow 0$, of lattice 
QCD does not meet with the naive expectation that this limit coincides 
with the heavy quark limit of real QCD~\cite{ste96}. 

\vskip 0.3cm 
The confining properties of Yang-Mills theory in 
the desired limit where the quark mass and the chemical potential are 
simultaneously made large (heavy quark limit) while the density of quarks 
is kept finite and non-zero were first investigated in~\cite{ben92,blu96}. 
Using Kogut-Susskind quarks, this limit simplifies the fermion determinant 
and allows a significant improvement of the statistics~\cite{blu96}. 
A recent breakthrough~\cite{kar99} was achieved by resorting 
to the heavy quark limit in addition to the use of the canonical 
ensemble, i.e.,  the hadron system of fixed density (by contrast to the 
grand-canonical ensemble of fixed chemical potential). 
The canonical approach involves the grand-canonical partition function 
with imaginary values of the chemical potential as first pointed out 
in~\cite{mil88}. A great simplification arises in this case from 
the fact that the fermion determinant is real. The 
numerical analysis of~\cite{kar99} reveals that the deconfinement 
phase transition becomes a cross-over at finite density (see 
also~\cite{blu96}). The string between static quark breaks yielding 
a constant heavy quark potential at large distances. 

\vskip 0.3cm 
In this paper, we present a new approach to Yang-Mills theory at a 
finite density of heavy quarks. We shall calculate the {\it continuum } 
fermion determinant for arbitrary entries for the gluon field 
in the large mass limit (rather than in the quenched approximation $N_f 
\rightarrow 0$) using the Schwinger proper-time regularization. The 
result is a gauge invariant action of the gluon fields, depending on 
an UV-regulator $\Lambda $, and is added to the standard action of 
Yang-Mills theory. The total result can be discretized on a lattice with 
lattice spacing $a$ by standard methods and can be used as input for computer 
simulations. In the critical limit $\Lambda \rightarrow \infty$, $a 
\rightarrow 0$, physical quantities become independent of the 
regularization scheme, and the results for observables are independent 
of the choice of the technique, i.e.,  lattice regularization or 
Schwinger proper-time regularization of the fermion determinant. 
The advantages of our 
approach are as follows: firstly, the starting point of the calculation 
is the continuum quark determinant with the correct number of degrees of 
freedom. The approach is not plagued by spurious states, and its 
heavy mass limit is manifestly the correct QCD limit. Secondly, despite 
the fact that 
the total gluonic action is still non-local, it is simple enough to allow 
for fast computer simulations. Thirdly, the correct scaling of physical 
quantities towards the continuum limit is manifestly the same as 
the one proposed by continuum Yang-Mills theory with quarks included. 

\vskip 0.3cm 
The paper is organized as follows: in the next section, we 
contrast the heavy quark limit with the quenched approximation, and 
discuss the difference in the renormalization group scaling in either 
case. The calculation of the (continuum) quark determinant in the large-mass 
limit is presented in section \ref{sec:3}. We address the renormalization 
of the coupled gluon quark system and discuss the approach of the 
continuum limit in lattice simulations of the joint system. 
At the end of section 3, we calculate the canonical partition function 
describing a system with definite baryon number. First numerical 
results for the case of an SU(2) gauge group are shown in section 4. 
The dependence of the quark density on the chemical potential for 
several values of the temperature below and above the deconfinement 
temperature is discussed. The conclusions are left to the final section.

\section{Heavy Fermions on the lattice } 
\label{sec:2} 

\subsection{ Setup } 

Our aim is to address the Yang-Mills theory at finite densities of 
baryons. For this purpose, the grand-canonical partition function 
in Euclidean space time, i.e.,  
\bea 
Z(\mu ) &=& \int {\cal D} q \; {\cal D} \bar{q} \; {\cal D} A_\mu \; 
\exp \left\{ - \int d^4x \; \left( L_Q + L_{YM} \right) \right\} \; , 
\label{eq:2.1} \\ 
L_{YM} &=& \frac{1}{4 g^2} \, F^{a}_{\mu \nu }[A] \, F^{a}_{\mu \nu }[A] 
\, 
\label{eq:2.2} \\ 
L_Q &=& \bar{q}(x) \; \left[ i \dslash \, - \, m \, - \, 
\gamma ^\mu A_\mu \, - \, i \mu \, \gamma ^0 \right] \; q(x) \; , 
\label{eq:2.3} 
\ena 
serves as a convenient starting point. Thereby, $F^a_{\mu \nu } [A]$ 
is the usual field strength tensor of the gluon field $A_\mu $, $g$ 
is the Yang-Mills gauge coupling strength, $\mu $ is the chemical 
potential and $m$ is the quark mass. We will assume $N_f$ quark 
flavors which are degenerate in mass. The convention for the 
$\gamma $-matrices can be found in appendix~A. A complete gauge fixing 
is understood in (\ref{eq:2.1}) with the gauge fixing terms included 
in the measure ${\cal D} A_\mu $. Below, we will employ the lattice 
version of the formulation (\ref{eq:2.1}--\ref{eq:2.3}) implying that 
we need not address the details of the gauge fixing. An alternative 
description of the grand-canonical partition function is obtained 
by integrating out the quark fields $q(x)$, $\bar{q}(x)$, i.e.,  
\bea 
Z(\mu ) &=& \int {\cal D} A_\mu \; \Det \left[ i \dslash \, - \, m 
\, - \, \gamma ^\mu A_\mu \, - \, i \mu \, \gamma ^0 \right] \; 
\exp \left\{ - \int d^4x \; L_{YM} \right\} \; , 
\label{eq:2.4} \\ 
&=& \int {\cal D} A_\mu \; \exp \left\{ - N_f \, S_{eff}[A](\Lambda, 
m, \mu ) \; - \; \int d^4x \; L_{YM} \right\} \; , 
\label{eq:2.5} 
\ena 
where $\Lambda $ is an ultra-violet regulator. Our goal will be 
to calculate $S_{eff}[A](\Lambda , m, \mu )$ for large values of 
the quark mass $m$. The result will be a gauge invariant functional 
of the gluon fields $A_\mu $. The joint action, $S_{eff} 
+ \int d^4x \; L_{YM}$, 
will then be discretized on a lattice of spacing $a$ and will be 
subject of computer simulations. In the critical limit, 
$a \rightarrow 0$, $\Lambda \rightarrow \infty $, physical observables 
will be independent of the type of regularization and will approach 
the continuum result.

\subsection{ Heavy fermion versus quenched limit } 

One can think of two limits for specifying the heavy quark 
approximation, i.e.,  
\bea 
\sigma ^{1/2} , T  & \ll & \Lambda \; \ll \; m \; 
\hbox to 5cm {\hfill (quenched limit) } 
\label{eq:2.6} \\ 
\sigma ^{1/2} , T   & \ll & m \; \ll \; \Lambda  \; 
\hbox to 5cm {\hfill (heavy quark limit) } \; , 
\label{eq:2.7} 
\ena 
where $T$ is temperature. The string tension $\sigma $ serves in this 
case as the typical energy scale of the pure Yang-Mills system. 
In the case of the quenched limit (\ref{eq:2.6}), the heavy quarks
are decoupled from the Yang-Mills system. One-loop perturbation 
theory (see e.g.~\cite{yn83}) then tells us that the continuum limit is 
approached via the scaling
\be 
M^2 \, a^2(\beta ) \; = \; \hbox{const.} \; \exp \left\{ \, - \, 
\frac{24 \pi ^2 }{ 11 N_c^2 } \, \beta \, \right\} \, , 
\label{eq:2.8} 
\en 
where $N_c$ is the number of colors, $\beta = 2N_c / g^2$, and $M$ is an 
arbitrary physical quantity of energy dimension one. 

\vskip 0.3cm 
By contrast, if we would like to associate the impact of charm, bottom 
and top quark on the gluonic sector with heavy quark physics, 
equation (\ref{eq:2.7}) must be considered as the physical heavy 
quark limit. In this case, these quark degrees of freedom contribute 
to the critical behavior, and one finds 
\be 
M^2 \, a^2(\beta ) \; = \; \hbox{const.} \; \exp \left\{ \, - \, 
\frac{24 \pi ^2 }{ N_c (11 N_c - 2 N_f) } \, \beta \, \right\} \, . 
\label{eq:2.9} 
\en 
A lattice Monte-Carlo simulation of SU($N_c$) gauge theory with 
$N_f$ quark flavors must recover the scaling (\ref{eq:2.9}) 
towards the continuum limit. In particular, this scaling must be 
obeyed in the physical heavy quark limit (\ref{eq:2.7}). 

\vskip 0.3cm 
Note that the quenched limit (\ref{eq:2.6}) is formally recovered 
from (\ref{eq:2.9}) by taking the limit $N_f \rightarrow 0$. 
The dependence of the bare coupling constant $g$ on the ultra-violet regulator 
$a$ provided by (\ref{eq:2.9}) must not be confused with the behavior of the 
renormalized coupling $g_R(s)$ on the renormalization point $s $, 
which for instance enters a renormalization flow analysis first 
proposed by Wilson~\cite{wil71}. In the latter case and for an  
energy scale $s < m_c$, where $m_c$ is the charm quark mass, 
the ''running'' of $g_R(s )$ with $s $ is dictated by the three 
light, so-called active, quark flavors~\cite{app75}.

\section{ Heavy fermions' action } 
\label{sec:3} 

\subsection{ The fermion determinant } 
\label{sec:3.1} 

The goal of this section is to calculate the fermion determinant 
\be 
D _F \; := \; \Det ^{N_f} \, \bigl( i\Dslash \, - \, 
i \mu \, \gamma ^0 \, - \, m \, \bigr) \; , \hbo 
D_\mu \, := \, \partial _\mu \; - \; A_\mu (x) \; , 
\label{eq:a1} 
\en 
resorting to a $1/m$ expansion where $m$ is the fermion mass. We will here 
only study the case where the masses of the quark flavors are equal. 
$N_f$ is the number of quark flavors and an UV-regularization is understood in 
(\ref{eq:a1}). $A_\mu (x)$ is the SU($N_c$) gauge field and $\mu $ the 
chemical potential. We will use anti-hermitian $\gamma $--matrices throughout 
this paper (see appendix~A). 

\vskip 0.3cm 
The determinant $D$ (\ref{eq:a1}) is a Lorentz scalar in four dimensions 
and therefore invariant under a reflection of all its vector entries, i.e.,  
$i \partial _\mu \rightarrow - i \partial _\mu$, $A _\mu \rightarrow - 
A _\mu$, $\mu \rightarrow - \mu $. Exploiting the anti-hermitian 
property of the $\gamma $-matrices, we therefore obtain 
\bea 
D _F &=& \Det ^{N_f/2} \, \left[  \left( i\Dslash \, - \, 
i \mu \, \gamma ^0 \, - \, m \, \right) \, \left( - i\Dslash \, + \, 
i \mu \, \gamma ^0 \, - \, m \, \right) \right] 
\nonumber \\ 
&=& \Det ^{N_f/2} \, \left[ \left( \Pi \, - \, i \mu \, 
\gamma ^0 \right) \; \left( \Pi ^\dagger \, + \, i \mu \, \gamma ^0 
\right) \right] \; , \hbo  \Pi \;  = \; i\Dslash \, - \, m \; . 
\label{eq:a2} 
\ena 
This equation shows the familiar result that the quark determinant is 
real for vanishing chemical potential. Eq.~(\ref{eq:a2}) also tells us 
that one generically expects an imaginary part of $D$ for real values 
of the chemical potential while the determinant $D$ is again real for 
purely imaginary entries of $\mu $ (note $ \gamma ^{0 \, \dagger } 
= - \gamma ^0 $). 

\vskip 0.3cm 
The functional determinant $D _F$ can be represented as a product of 
eigenvalues, i.e.,  
\bea 
D _F &=& \left[ \prod ^{\{reg\}} _n \lambda _n \right] ^{N_f/2} \; 
\label{eq:al1} \\ 
\left[ \left( \Pi \, - \, i \mu \, 
\gamma ^0 \right) \; \left( \Pi ^\dagger \, + \, i \mu \, \gamma ^0 
\right) \right] \; \psi _n (x) &=& \lambda _n \; \psi _n (x) \; , 
\label{eq:al2} \\ 
\psi _n (x^0+1/T,\vec{x}) &=& (-1) \, \psi _n (x^0,\vec{x})) \; ,
\label{eq:al3} 
\ena 
It is convenient for technical reasons (see also~\cite{kar99}) to remove 
the chemical potential $\mu $ from the operator by a scale transformation 
of the spinor $\psi $, i.e.,  
\bea 
\psi &\rightarrow & \psi ^\prime (x) \; = \;  \exp \{ - \mu x^0 \} \psi (x) 
\; , 
\nonumber \\ 
\Pi \, \Pi ^\dagger \; \psi^\prime  _n (x) &=& \lambda _n \; \psi^\prime  
_n (x) \; , 
\label{eq:a4} \\ 
\psi^\prime _n (x^0+1/T,\vec{x}) &=& \left[ - \exp (- \mu /T ) \right]  
\, \psi ^\prime _n (x^0,\vec{x})) \; . 
\label{eq:a5} 
\ena 
Introducing the gauge covariant differential $D_\mu := \partial _\mu 
+ i A_\mu $ and the field strength $F_{\mu \nu }$, i.e.,  
\be 
F_{\mu \nu } \; = \; -i \, [D_\mu, D_\nu ] \; , 
\label{eq:3.3} 
\en 
we then find 
\be 
\Pi \, \Pi ^\dagger  \; = \;  - D_\mu D_\mu \, + \, m^2  \, + \, 
\frac{ \sigma _{\mu \nu } }{2} F_{\mu \nu } \; . 
\label{eq:3.4} 
\en 
Schwinger's proper-time method provides a gauge invariant 
regularization of functional determinants. In particular, the contribution 
of the fermion determinant $D$ to the gluonic action becomes 
\be 
- \, \ln \, D \; = \; 
\frac{ N_f }{2} \, \int d^4x \; \lim _{x \to y } \, 
\, \int _{1/\Lambda ^2 }^{\infty } \frac{ d \tau }{ \tau } \; 
\tr \, K(\tau ; x,y) \; . 
\label{eq:3.5} 
\en 
The so-called heat kernel $K(\tau ; x,y)$ satisfies the equation 
\be 
\frac{ \partial }{ \partial \tau } K(\tau ; x,y) \; + \; 
\Pi \, \Pi ^\dagger \, K(\tau; x,y ) \; = \; 0 
\label{eq:3.6} 
\en 
and the boundary conditions $(t \equiv x^0)$
\bea 
K(\tau =0; x,y) &=& \delta (x-y) \; , \hbox to 3cm {\hfill for \hfill } 
x,y \in V \; , 
\label{eq:3.7} \\ 
K[\tau; (t_x  + \frac{n}{T}, \vec{x}) , (t_y + \frac{m}{T}, 
\vec{y})] &=& (-e^{- \mu/T})^{n-m} \; K(\tau; x, y) \; , 
\label{eq:3.8} \\ 
\lim_{\vec{x}, \vec{y} \to \infty } K(\tau; x, y) &=& 0 \; , 
\label{eq:3.9} 
\ena 
where $V$ is the space time volume of the (lattice) universe, $T$ is the 
temperature and $n, m \in \mathbbm{Z}$. At the present stage, 
$V$ is considered to be a cylinder with a periodicity of $1/T$ in 
time direction and infinite extension in spatial directions. 
Although the source 
$\Pi \, \Pi ^\dagger $ which enters the heat equation (\ref{eq:3.6}) 
is hermitian, the desired imaginary parts of (\ref{eq:3.5}) will 
originate (see below) from the non-trivial topology of the non-simply  
connected space-time manifold~\cite{holger}. 

\vskip 0.3cm 
In order to derive the systematic expansion in powers of the 
inverse fermion mass, i.e.,  $1/m$, we shall adopt the heat-kernel 
expansion, and resort to the techniques reported in the important paper by 
Ebert and Reinhardt~\cite{eb86}. This expansion is generated by the ansatz 
\be 
K(\tau; x,y) \; = \; K_0(\tau; x,y) \, H(\tau; x,y) \; , \hbo 
H(\tau; x,y) \; = \; \sum _{j=0}^\infty h_j (x,y) \tau ^j \; ; 
\label{eq:3.14} 
\en 
By definition $K_0(\tau; x,y)$ satisfies the equation 
\be 
\frac{\partial }{ \partial \tau } \, K_0(\tau; x,y) \; + \; 
\left[ - \partial ^2 \, + \, m ^2 \, \right] \, 
K_0(\tau; x,y) \; = \; 0 \; . 
\label{eq:3.15} 
\en 
A particular solution to this equation is given by~\cite{eb86} 
\bea 
K_0(\tau; x,y) &=& \frac{1}{16 \pi ^2 \tau ^2 } \; 
\label{eq:3.16} \\ 
&& \exp \left\{ -m^2 \tau \, - \, \frac{ (t_x-t_y + \alpha )^2 + 
(\vec{x}-\vec{y})^2 }{ 4 \tau } \right\} \; , 
\nonumber 
\ena 
where $\alpha $ is an arbitrary constant. The interacting part, 
$H(\tau ; x,y)$, of the heat kernel satisfies the equation 
\be 
\left( \frac{ \partial }{ \partial \tau }  \; + \; \frac{1}{\tau } 
\, z_\mu D_\mu \; - \; D_\mu D_\mu \; + \; \frac{1}{2} \sigma _{\mu \nu } 
F_{\mu \nu } \; \right) \; H(\tau ; x,y) \; = \; 0 \; , 
\label{eq:3.17} 
\en 
where $z_\mu = x_\mu - y_\mu + \alpha \delta _{\mu 0}$. Suppose that 
$H(\tau; x,y)$ provides a solution to equation (\ref{eq:3.17}). 
Since the gluon field satisfies periodic boundary conditions, i.e.,  
\be 
A_\mu (t + n/T, \vec{x} ) \; = \; A_\mu (t , \vec{x} ) \; , 
\label{eq:3.18} 
\en 
this solution possesses a discrete translation invariance, i.e.,  
\be 
H\Big( \tau ; (t_x + n/T, \vec{x}), (t_y + n/T, \vec{y}) \Big)
\; = \; H\Big( \tau ; (t_x , \vec{x}), (t_y , \vec{y}) \Big) \; , 
\hbo n \; \hbox{ integer } \; . 
\label{eq:3.18a} 
\en 
A close inspection of (\ref{eq:3.17}) then shows that 
$$ 
H\Big( \tau ; (t_x + n/T, \vec{x}), (t_y , \vec{y}) \Big)
$$ 
is also a solution of (\ref{eq:3.17}) if one chooses $\alpha =n/T$. 
Equipped with these prerequisites 
we arrive at the central result of this section: the heat kernel which 
satisfies the boundary conditions (\ref{eq:3.8}) is given by 
\bea 
K(\tau; x,y) &=& 
\sum _n \, (-e^{ \mu /T} )^n \; \frac{1}{16 \pi ^2 \tau ^2 } 
\label{eq:3.20} \\ 
&& \exp \left\{ -m^2 \tau \, - \, \frac{ (t_x-t_y + n/T)^2 + 
(\vec{x}-\vec{y})^2}{ 4 \tau } \right\} 
\nonumber \\ 
&& H\Big( \tau ; (t_x + n/T, \vec{x}), y) \Big) 
\; . 
\nonumber 
\ena 
The so-called diagonal parts of the heat--coefficients, i.e.,  
$\lim _{x \to y } h_j(x,y)$, can be found in~\cite{eb86}. 
In order for ensuring proper boundary conditions imposed by finite 
temperatures, the full functional dependence of $h_j (x,y)$ on $x,y$ 
is needed. The calculation of the complete heat--coefficients 
for $j =0,1,2$ is one of our goals in the present paper. The 
explicit calculation is shown in appendix~B. Below, we will only 
make use of $h_0 (x,y)$ which is given by 
\be 
h_0 (x,y) \; = \; {\cal P } \, \exp \left\{ -i \int _{C_{xy}} 
A_\mu (x^\prime ) \; dx_\mu ^\prime \, \right\} \; , 
\label{eq:3.21} 
\en 
where $C_{xy}$ is a straight line connecting the points $x$ and $y$. 
The result for $h_1 (x,y)$ and $h_2 (x,y)$ can be found in 
appendix~B. 
Inserting (\ref{eq:3.20}) into (\ref{eq:3.5}) while employing the 
expansion (\ref{eq:3.14}) generates the $1/m$ expansion 
of the fermionic contribution to the gluonic action, i.e.,  
\bea 
S_{f} &=&  - \, \ln \, D _F \; = \; 
N_f \, \sum _n \, \sum _{j =0 } \; \left[ - e^{\mu /T } \right]^n
\label{eq:3.25} \\ 
&& \frac{1}{8 \pi ^2 } \, \int _{1/\Lambda ^2 }^\infty \frac{ d\tau }
{ \tau ^{3-j } } \; \exp \left\{ - m^2 \tau - 
\frac{ n^2 }{ 4 T^2 \tau } \right\} \; \; 
\int d^4x \; \tr h_j \left( \left(t_x + n/T, \vec{x} \right), x 
\right) \; . 
\nonumber 
\ena 
For the later discussion, it is convenient to single out the 
temperature independent part $n=0$. Performing the $\tau $-integration 
we finally obtain 
\bea 
&& S_f \; = \;  \frac{N_f}{8 \pi ^2} \; \sum _{j = 0 } \,\left\{ 
m^{4-2j } \; \Gamma \left( j -2, \frac{m^2}{\Lambda ^2 } \right) 
\; \int d^4x \; \tr h_j (x,x) 
\right. \label{eq:3.26} \\ 
&+& \left. \frac{N_f}{2} \sum _{n \not= 0 } \left[ - e^{\mu /T } 
\right]^n \, m^{4-2j } \, I_j \left(n^2 \frac{m^2}{4 T^2 }, 
\frac{m^2}{\Lambda ^2} \right) \, \int d^4x \, \tr h_j \left( \left(t_x 
+ n/T, \vec{x} \right), x \right) \right\}\, , 
\label{eq:3.27} 
\ena 
where the trace extends over color only, and where 
$\Gamma ( j, x)$ is the incomplete gamma function, 
\be 
\Gamma ( j, x) \; = \; \int _{x}^\infty ds \; s^{j -1} \, e^{-s} \; , 
\label{eq:3.28} 
\en 
and where 
\be 
I_j (x,y) \; = \; \frac{1}{4 \pi ^2} \; \int _y^\infty 
\frac{ ds}{s^{3-j }} \; \exp \left\{ - s \, - \, \frac{x}{s} 
\right\} \; . 
\label{eq:3.29} 
\en 
Since the limit $\lim _{y \to 0} I_j (x,y)$ does exist, only the 
term $n=0$ (\ref{eq:3.26}) picks up a divergence. 
This observation reflects the familiar fact that only temperature 
independent terms are affected by ultra-violet divergences.

\subsection{ Renormalization } 

In this subsection, we will study the UV-divergence which emerge 
in (\ref{eq:3.26}). Since in this equation only the diagonal 
part of the heat coefficients are involved, one might resort to 
the derivation of the diagonal parts presented by Ebert and Reinhardt 
in~\cite{eb86}. One finds (see also appendix B) 
\be 
\tr h_0 \; = \; N_c \; , \hbo \tr h_1 \; = \; 0 \; , \hbo 
\tr h_2 (x,x) \; = \; \frac{1}{6} \, \tr F_{\mu \nu } F_{\mu \nu } 
\; . 
\label{eq:3.50} 
\en 
The only non-trivial gluon field dependence is therefore induced 
by the $h_j $, $j \ge 2$ terms. Since the limit $\lim _{x \to 0 } 
\Gamma (j-2, x)$ exists for $j \ge 3$, only the term of 
(\ref{eq:3.26}) with $j =2$ develops a singularity. In fact, one 
obtains for a large cutoff $\Lambda $ 
\be 
\Gamma \left(0, \frac{m^2}{\Lambda ^2} \right) \; = \; 
\ln \; \frac{ \Lambda ^2 }{ m^2 } \; + \; {\cal O }(1) \; . 
\label{eq:3.51} 
\en 
The divergent part of the fermionic action can therefore be calculated 
analytically, 
\be 
S_{f}^{div} \; = \; \frac{N_f}{32 \pi ^2} \int d^4x \; \ln \frac{ 
\Lambda ^2 }{ m^2 } \; \frac{1}{3} F^a_{\mu \nu } F^a_{\mu \nu } \; . 
\label{eq:3.52} 
\en 
Renormalization is accomplished by absorbing the divergences by 
an appropriate choice of the bare coupling strength $g$, i.e.,  
\be 
\lim _{\Lambda \to \infty } \; \left[ \frac{1}{4 g^2} \int d^4x \; 
F^a_{\mu \nu } F^a_{\mu \nu } \; + \; S^{div}_{gluon} \; + \; S_f^{div} 
\right] \rightarrow \hbox{finite} \; . 
\label{eq:3.53} 
\en 
One loop perturbation theory of Yang-Mills theory {\it without } quarks 
yields 
\be 
\frac{1}{4 g^2 } \; - \; \frac{N_c}{32 \pi ^2 } \frac{11}{6} \, 
\ln \frac{ \Lambda ^2 }{ \mu ^2 } \rightarrow \hbox{finite} \; , 
\label{eq:3.54} 
\en 
where $N_c$ is the number of colors and 
$\mu $ is an arbitrary renormalization scale. {\it Including } 
$N_f$ quark flavors, we make use of (\ref{eq:3.52}) and demand 
\be 
\frac{1}{4 g^2 } \; - \; \frac{N_c}{32 \pi ^2 } \frac{11}{6} \, 
\ln \frac{ \Lambda ^2 }{ \mu ^2 } 
\; + \; \frac{N_f}{32 \pi ^2 } \frac{1}{3} \, 
\ln \frac{ \Lambda ^2 }{ m ^2 } \rightarrow \hbox{finite} \; . 
\label{eq:3.55} 
\en 
The renormalization group $\beta $-function, 
\be 
\beta (g) \; := \; \frac{1}{g} \frac{ dg (\Lambda ) }{d \ln \Lambda } \; ,
\label{eq:3.56} 
\en 
can be readily calculated from (\ref{eq:3.55}). We recover the 
well known result 
\be 
\beta (g) \; = \; - \; \frac{g^2}{8 \pi ^2} \, \left( 
\frac{11N_c}{6} \, - \, \frac{N_f}{3} \right) \; + \; {\cal O}(g^3) \; . 
\label{eq:3.57} 
\en 
The only role of the heavy fermion determinant at this level 
of the large $m$-expansion is to generalize the renormalization 
group $\beta $-function of the pure Yang-Mills theory to the situation 
with $N_f$ quark flavors included. 

\vskip 0.3cm 
We are now in the position to answer the question which raises from  
subsection~2.2: how can the correct scaling towards the continuum limit 
be observed in lattice Yang-Mills theory with $N_f$ quarks included? 
If we neglect for the moment the finite-temperature corrections 
(\ref{eq:3.27}), the inclusion of fermions only affects the 
parameter $\beta = 2N_c/g^2 $ in front of the Wilson action, i.e.,  
\be 
\beta \; \rightarrow \; \beta _F \; = \; \beta \, - \, 
\frac{N_f}{2 \pi ^2 } \, \frac{N_c}{6} \, \ln (a^2 \sigma ) \; , 
\label{eq:3.58} 
\en 
where we have used the freedom in choosing the (fermionic) cutoff 
for defining 
\be 
\frac{\Lambda ^2 }{m ^2 } \; = \; \frac{1}{ \sigma a^2} \; , 
\label{eq:3.59} 
\en 
where $a$ is the lattice spacing. $\sigma $ is the string tension
of pure, i.e.,  $N_f=0$, Yang-Mills theory and serves as reference scale. 
Let us assume that we calculate some physical mass $M$ in units of the 
lattice spacing as function of the only parameter $\beta _F$. 
Since the numerical simulation is carried out with the standard 
Wilson action (with a coefficient changed from $\beta $ to $\beta _F$), 
we recover the familiar scaling behavior at large values 
of $\beta $ which is at one loop level 
\be 
M^2 \, a^2(\beta _F) \; = \; \kappa \, \exp \left\{ - 
\frac{ 24 \pi ^2 }{11 \, N_c^2} \, \beta _F \, \right\} \; , 
\label{eq:3.60} 
\en 
where $\kappa $ is a numerical constant which must be ''measured'' 
by the numerical simulation. Inserting $\beta _F$ (\ref{eq:3.58}) 
in (\ref{eq:3.60}) elementary manipulations of this equation finally yield 
\be 
M^2 \, a^2 \; = \; \kappa \, \left( \frac{ \kappa \sigma ^2 }{M^2} 
\right) ^{\frac{ 2 N_f}{ 11 N_c- 2N_f } } \; \exp \left\{ 
- \frac{24 \pi ^2 }{ N_c (11 N_c - 2 N_f } \, \beta \right\} \; . 
\label{eq:3.61} 
\en 
The crucial observation is that $M^2 \, a^2 $ exponentially decreases 
with $\beta $ while the slope precisely meets with the expectations for 
a SU($N_c$) gauge theory with $N_f$ quark flavors. The result of 
this subsection is also of practical importance: performing 
numerical simulations with $\beta _F$ as coefficient of the 
Wilson action and observing the standard scaling (\ref{eq:3.60}) 
automatically implies the correct approach of the continuum limit 
with $N_f$ fermions included. 

\vskip 0.3cm 
Let us assume that we have calculated two physical observables 
$M_1$ and $M_2$ and that we have obtained the proper $\beta $-dependence 
(\ref{eq:3.61}) for either quantity at asymptotic values of $\beta $. 
The ratio of both observables then becomes 
$$ 
\frac{ M_1^2 }{ M_2^2 } \; = \; \frac{ \kappa _1 }{ \kappa _2 } 
\, \left( \frac{ \kappa _1 M_2^2 }{ \kappa _2 M_1^2 } 
\right) ^{\frac{ 2 N_f}{ 11 N_c- 2N_f } } . 
$$ 
Using e.g.~$M_2$ as reference scale one expresses $M_1$ in terms 
of the ''measured'' quantities $\kappa _1 $ and $\kappa _2$, i.e., 
$$  
M_1 \; = \; \frac{ \kappa _1 }{ \kappa _2 } \, M_2 \; . 
$$ 

We finally note that temperature dependent terms (\ref{eq:3.27}), which 
we have not considered in this subsection, are finite in the 
limit $a \rightarrow 0$. We therefore do not expect that these terms will 
change the critical behavior for $\beta \rightarrow \infty $ 
in (\ref{eq:3.61}).

\subsection{ Lattice YM-theory with finite chemical potential } 
\label{sec:3.3} 

In this subsection, we will consider the temperature dependent part 
of the sum (\ref{eq:3.27}). The pre-factors of the heat coefficients
$h_j ( (t_x + n/t, \vec{x}), x) $ are given by the functions 
$I_j (n^2 m^2/4 T^2, m^2/\Lambda ^2)$ which stay finite when the 
UV regulator is removed $(\Lambda \rightarrow \infty )$. Using the heavy 
quark, but non-quenched limit $m^2/\Lambda ^2 \rightarrow 0 $ 
(see discussion in subsection~2.2), one finds 
\bea 
m^{4- 2 j } \, I_j \left(n^2 \frac{m^2}{4 T^2 }, 0 \right) &=& 
\frac{ m^{4- 2 j } }{4 \pi^2} \; \int _{0}^{\infty } 
\frac{ ds}{s^{3-j } } \; \exp \left\{ - s \; - \; 
\frac{ n^2 m^2  }{ 4 T^2 s } \right\} 
\label{eq:3.70} \\ 
&=& \frac{ 2 m^2 T^2 }{ \pi ^2 \, n^2 } \;\frac{ n^j }{ (2 m T)^j } 
\; K_{2-j }  \left( \frac{ n m }{T} \right) \; , 
\nonumber 
\ena 
where the $K_n(x)$ are the modified Bessel functions of the second kind. 
The heat coefficients $h_j $ are functionals of the gluonic fields 
only and carry energy dimension $2j $. One therefore expects that 
their order of magnitude is given by $\Lambda ^{2j } _{YM}$ 
where $\Lambda  _{YM}$ is the characteristic energy scale of pure 
Yang-Mills theory. This implies that the sum over $j $ in 
(\ref{eq:3.70}) generates the desired heavy quark expansion in powers of 
$\Lambda ^{2} _{YM} / m T $. In the following, we will assume that 
the mass is large enough for a truncation of this sum at $j =0 $.

Using (\ref{eq:3.21}) one might interpret 
\be 
h_0 \left( \left(t_x + 1/T, \vec{x} \right) \right) \; =: \; P(x) 
\label{eq:3.71} 
\en 
as Polyakov line $P(x)$ which starts at the space time point $x$ and which 
winds around the torus in time direction finally ending at the same 
point $x$. The operator $P(x)$ 
can be directly translated to the corresponding lattice operator: 
it is the (path-ordered) product of the link variables $U_\mu (x)$ 
along the $\mu =4 $ direction. Expressing the field strength 
squared in terms of the plaquette variable $P_{\mu \nu }(x)$, i.e.,  
\be 
\frac{1}{4} \, F^c_{\mu \nu } \, F^c_{\mu \nu } \; a^4 \; = \; 
N_c \; \biggl[ 1 \; - \; \frac{1}{N_c} \, \tr P_{\mu \nu }(x) \biggr] \; , 
\label{eq:3.71a} 
\en 
we finally obtain the lattice action $S_{latt}$ of SU(2) Yang-Mills theory 
with $N_f$ heavy quark flavors 
\bea 
-S_{latt} &=& - \; S_F \; + \; \beta _F \, \sum _{(\mu > \nu) \{x\} } 
P_{\mu \nu }(x) 
\label{eq:3.72} \\ 
-S_F &=& - \; \sum _{\{x\}} \biggl( \; \frac{ m^2 T^2 }{\pi ^2} a^4 \, 
\sum _{n=1}^\infty \, \frac{ (-1)^n }{n^2}  K_2 \left( \frac{nm}{T} \right) \, 
\label{eq:3.72a} \\ 
& \phantom{-} & \phantom{ \sum _{\{x\}} \biggl(  }
\left[ e^{ n \frac{\mu }{T} } \; \tr P^n(x) \; + \; 
e^{ - n \frac{\mu }{T} } \; \tr \left( P^\dagger (x)^n \right) 
\right] \; \biggr) \; . 
\nonumber 
\ena 
where $\beta _F$ is given by (\ref{eq:3.58}). Exploiting the heavy-quark 
limit, i.e.,  $m \gg T$, one uses the asymptotic expression for 
Bessel functions, 
\be 
K_2(x) \; = \; \sqrt{ \frac{ \pi }{ 2x } } \; e^{-x} \; 
\left[ 1 \, + \, {\cal O} \left( \frac{1}{x} \right) \, \right] \; . 
\label{eq:3.73} 
\en 
In this limit $-S_F$ in (\ref{eq:3.72a}) becomes 
\be 
- S_F \; = \; - \sum _{\{x\}}
\frac{ m^2 T^2 }{\pi ^{3/2} } a^4 \, \sqrt{ \frac{T}{2m} } \, 
\sum _{n=1}^\infty \, \frac{ (-1)^n }{n^{5/2} }  \, 
\left[ e^{ n \frac{\mu - m }{T} } \; \tr P^n(x) \; + \; 
e^{ - n \frac{\mu + m }{T} } \; \tr \left( P^\dagger (x)^n \right)  
\right]  \; . 
\label{eq:3.73a} 
\en 
For a SU(2) gauge group, this equation is real even at finite 
values of the chemical potential as it is $\tr P $. By contrast, 
one expects imaginary parts for real $\mu $ and SU($N_c\ge 3)$. 
Finally, one observes that for purely imaginary entries $\mu = i \nu $, 
$\nu $ real, equation (\ref{eq:3.73a}) is real as expected (see 
discussion in subsection \ref{sec:3.1}). 

\vskip 0.3cm 
For $\mu \le m $, the sum in (\ref{eq:3.73a}) is rapidly converging, 
while the sum is only asymptotic for $\mu \gg m $ 
(for a discussion of this issue in the context of QED 
see~\cite{elm94}). 
In the later case, resorting to different representations of the 
Bessel function $K_2$ it is possible to perform an analytic 
continuation from the finite sum for $\mu \le m $ to large values 
of $\mu >m $~\cite{hol99}. The outcome of this procedure is that 
by means of analytic continuation one can assign a finite and real 
value to the sum also for $\mu > m$. The lack of an imaginary part 
possesses a physical interpretation: 
the chemical potential $\mu $ describes the 
gain in energy if a particle is added to the system, while an energy $E$ 
is necessary to produce the particle. 
Neglecting binding and confinement effects (for this argument only), 
the particle must occupy an empty phase space cell and carries a 
momentum larger than the Fermi momentum. The loss in energy due to 
particle production is therefore always larger than the gain $\mu $ 
in energy. The system is stable. 

\vskip 0.3cm 
The production of 
quarks at $\mu > m $ is conflicting with quark confinement, which does 
not tolerate single quark productions. This interplay between these 
mechanisms can be anticipated from (\ref{eq:3.73a}). One expects 
a drastic influence of the heavy quark corrections to the 
zero density action if 
\be 
\exp \left\{ \frac{ \mu - m }{T} \right\} \; \langle \vert P \vert 
\rangle \; > \; 1 \; . 
\label{eq:3.74} 
\en 
For small temperatures (and moderate densities), $ \langle 
\vert P \vert \rangle $ is small, and the onset of the density effects 
is postponed to values $\mu \gg m $. On the other hand, at high 
temperatures $(T> \Lambda _{YM})$, 
$ \langle \vert P \vert \rangle$ is of order one by 
temperatures effects only, and a significant rise of density is 
expected for $\mu \approx m $. From a physical point of view, 
this rise of density might be interpreted as single quark productions which 
become feasible in the deconfinement region. 

\vskip 0.3cm 
In the following, we will study the case where $\mu $ is only slightly 
larger than $m$. This procedure will allow us to study baryon 
matter at moderate densities. We hope that in this case the truncation of the 
sum at $n=1$ already captures the essence of the asymptotic series 
(\ref{eq:3.73a}). The extension of the considerations to larger values 
of the chemical potential is an interesting task which is left to 
future investigations. In the present paper, we will only perform a 
consistency check by estimating the $n=2$ term. Within this approximation, 
we finally obtain 
\bea 
-S_{latt} &=& \sum _{\{x\}} \biggl( \beta _F \, \sum _{\mu > \nu } 
P_{\mu \nu }(x) 
\label{eq:3.75} \\ 
&+& \frac{ m^{3/2} T^{5/2} }{\sqrt{2} \pi ^{3/2} } a^4 \, 
\left[ e^{ \frac{\mu - m }{T} } \; \tr P(x) \; + \; 
e^{ - \frac{\mu + m }{T} } \; \tr P^\dagger (x) \, 
\right] \; \biggr) \; . 
\nonumber 
\ena 
The partition function which describes the interaction of SU(2) 
gauge fields with $N_f$ heavy fermions is finally expressed as
an integral over the link variables $U_\mu (x)$, i.e.,  
\be 
Z(\mu ) \; = \; \int {\cal D} U_\mu (x) \; \exp \{ - S_{latt} 
\} \; . 
\label{eq:3.76} 
\en 
Equations (\ref{eq:3.75},\ref{eq:3.76}) are the main results of the 
present paper. 
Note that the considerations of heavy fermions induce non-local 
terms to the effective action of gluons. These terms are given 
by the Polyakov line in (\ref{eq:3.75}). This non-locality is confined 
to the temporal direction while the action is still local in 
spatial directions. Note further that $\tr P(x) $ is real for a 
SU(2) gauge group. In the SU(2) case the action in (\ref{eq:3.72}) can be 
therefore easily simulated with a moderate increase in 
computer time compared with the case of pure YM theory by resorting to the 
standard heat-bath algorithm. First results of such a simulation will 
be presented below. 
 
\vskip 0.3cm 
Let us finally calculate the baryon density $\rho $ from the functional 
integral (\ref{eq:3.76}). The derivative of the partition function 
with respect to the chemical potential yields 
\be 
\frac{ d \, \ln Z(\mu )}{ d\mu } \; = \; \frac{N_c B}{T} \; , 
\label{eq:3.77} 
\en 
where $B$ is the number of Baryons which are present in the 
(lattice) universe. Inserting (\ref{eq:3.76}) and 
(\ref{eq:3.75}) into (\ref{eq:3.77}), one finds for the baryon density 
\be 
\rho \; = \; B/V \; = \; 
\frac{ (m T)^{3/2} }{N_c \sqrt{2} \pi ^{3/2} } \, 
\left[ e^{ \frac{\mu - m }{T} } \; \langle \tr P \rangle \; - \; 
e^{ - \frac{\mu + m }{T} } \; \langle \tr P^\dagger \rangle \, 
\right] \; , 
\label{eq:3.78} 
\en 
where $V= N^3 _\sigma a^3$ is the volume and where $T=1/N_t a$ 
was used. $N_\sigma $ and $N_t$ denote the number of lattice point 
in spatial and time direction, respectively. 
$\langle \tr P \rangle $ is the expectation value of the 
Polyakov line. For positive values of the chemical potential and $\mu 
\approx m $, thermal excitations of anti-quarks can be neglected. 
In this case, the second term on the right hand side of (\ref{eq:3.78}) 
can be neglected, and the density is triggered by the expectation 
value of the Polyakov loop.

\subsection{ Lattice YM-theory at fixed baryon density } 
\label{sec:3.4} 

Endowed with the results of the previous subsection, it is an easy task 
to apply the approach of Miller and Redlich~\cite{mil88} for 
describing the Yang-Mills theory with a definite value of the baryon 
density. Their pioneering approach is based on the observation that 
the partition function 
\be 
Z_\rho (B) \; = \; N_c \; \int \frac{ d \nu }{ 2 \pi T } \; 
\exp \left\{ - i \nu \frac{ N_c B}{T} \right\} \; Z(i \nu ) 
\label{eq:3.80} 
\en 
describes the Yang-Mills theory at a fixed density provided by 
$B$ baryons in the universe. Thereby, $Z(i \nu )$ is the 
grand-canonical partition function (\ref{eq:2.1}) with an imaginary 
entry as chemical potential. Inserting (\ref{eq:2.1}) into 
(\ref{eq:3.80}), one readily verifies that the $\nu $-integration 
constrains the parameter $B$ to the baryon number, i.e.,  
\be 
B \; = \; \frac{1}{N_c} \, \int d^3x \; \biggl\langle \, i \, 
\bar{q}(x) \gamma ^0 q(x) \, \biggr\rangle _t \; , 
\label{eq:3.81} 
\en 
where an average over a time period of length $1/T$ is understood 
in (\ref{eq:3.81}). Note that $T$ denotes the temperature, and that 
we use anti-hermitian $\gamma $-matrices (see appendix A). 

\vskip 0.3cm 
One easily repeats the derivation of the heavy quark lattice action 
(\ref{eq:3.72}) for the case of an imaginary chemical potential. 
The calculation of the eigenmodes of the heat kernel (subsection 
\ref{sec:3.1}) is unaffected by the choice of an imaginary chemical 
potential, and one essentially observes a change of the boundary 
conditions (\ref{eq:a5}) of the quark fields. The final result 
of this consideration is that the heavy quark lattice action 
(\ref{eq:3.72}) is given in the case of an imaginary chemical 
potential by replacing $\mu \rightarrow i \nu $ in (\ref{eq:3.72}). 

\vskip 0.3cm 
In accordance with~\cite{blu96,kar99}, one immediately realizes that the 
heavy quark lattice action (\ref{eq:3.73a}) is real for an imaginary chemical 
potential. In agreement with~\cite{kar99}, we also observe that 
the partition function (\ref{eq:3.80}) for a given baryon number $B$ 
is invariant under a center transformation of the links which belong 
to a spatial hypercube at a given time slice. In this case, the 
Polyakov loop acquires a phase $\exp (i\,  2\pi /N_c ) $ 
which, however, can be absorbed by a redefinition of the integration 
variable $\nu \rightarrow \nu - 2\pi / N_c $.

\vskip 0.3cm 
By contrast to the case of a finite real value of the chemical potential, 
the sum over $n$ in~(\ref{eq:3.73a}) is rapidly converging for 
$\mu = i \nu $ as long as $m > T$. Calculating the fermionic 
contribution $\exp \{ - S_F \}$ to the probabilistic weight from 
(\ref{eq:3.73a}), one observes that the terms $P^n(x)$ with a multiple, i.e.,  
$n>1$, winding of the Polyakov loop around the torus are suppressed 
by the factor $1/n^{5/2}$ in~(\ref{eq:3.73a}). One finds 
\bea 
e^{-S_F} & \approx & \prod _{\{x\}} \left\{ 1 \; + \; \zeta 
\left[ e^{ i \nu /T  } \; \tr P(x) \; + \; 
e^{ - i \nu /T } \; \tr P^\dagger (x) \right]  \right\} \; , 
\label{eq:3.90} \\ 
\zeta &=& \frac{ m^2 T^2 }{\pi ^{3/2} } a^4 \, \sqrt{ \frac{T}{2m} } \, 
e^{-m/T} \; . 
\label{eq:3.91} 
\ena 
Inserting (\ref{eq:3.90}) into (\ref{eq:3.80}), the Fourier integration 
can be explicitly performed. The final result is 
\be 
Z_\rho (B) \; \approx \; \biggl[ \zeta ^{N_c B} \, \sum _{\{x\}} 
\prod _{i=1}^{N_c B} \tr P(x_i) \, \biggr] \; \exp 
\sum _{\{x\}} \biggl( \beta _F \, \sum _{\mu > \nu } 
P_{\mu \nu }(x) \biggr) \; , 
\label{eq:3.100} 
\en 
where $x_i \not= x_k $ for $i\not=k$ holds. Insertions of 
$P (x_a) P^\dagger (x_b)$ are suppressed by a factor $\zeta ^2 $ 
and do not contribute to the leading order result (\ref{eq:3.100}). 
From a physical point of view, these insertions correspond to 
particle anti-particle excitations which are suppressed by a factor 
of roughly $\exp \{- 2m /T \}$ in the probabilistic weight. Note 
that $ Z_\rho (B)$ is manifestly center symmetric since only 
products of Polyakov loops which consist of a multiple of $N_c$ 
factors are present in (\ref{eq:3.100}). 

\vskip 0.3cm 
It is instructive to compare the partition function $ Z_\rho (B) $ 
of a system of finite baryon density, i.e.,  $B \not= 0$, with 
the zero density result $ Z_\rho (B=0) $. One observes that the 
finite density partition function is suppressed by a factor 
$\zeta ^{N_c B}$ relative to the zero density case. This illustrates 
the overlap problem which rises by resorting to the 
Glasgow algorithm~\cite{bar99}. Simulating the partition function 
of a system with $B$ baryons employing an algorithm with 
generates an Monte-Carlo ensemble which is based on the zero density 
partition function requires a huge amount of statistics to 
overcome the entropy factor $\zeta ^{N_cB}$. 

\vskip 0.3cm 
It was argued by Svetitsky~\cite{sve86} quite some time ago that 
the Polyakov line corresponds to a field configuration with an 
essential overlap with the wave function of a static quark. At that time, 
this approach paved the way for the interpretation of the 
Polyakov line expectation value as an order parameter for the 
deconfinement phase transition. Our result (\ref{eq:3.100}) 
nicely confirms Svetitsky's considerations on a quantitative level. 

\vskip 0.3cm 
We finally point out that in a recent publication~\cite{for99} a close 
relation of the phase of the (lattice) fermion determinant and the 
imaginary parts of the Polyakov loop was observed. 
Our results confirm these findings.

\section{ Numerical simulations} 
\label{sec:4} 

\begin{figure}[t]
\parbox{9cm}{ 
\hspace{1cm} 
\centerline{ 
\epsfxsize=9cm
\epsffile{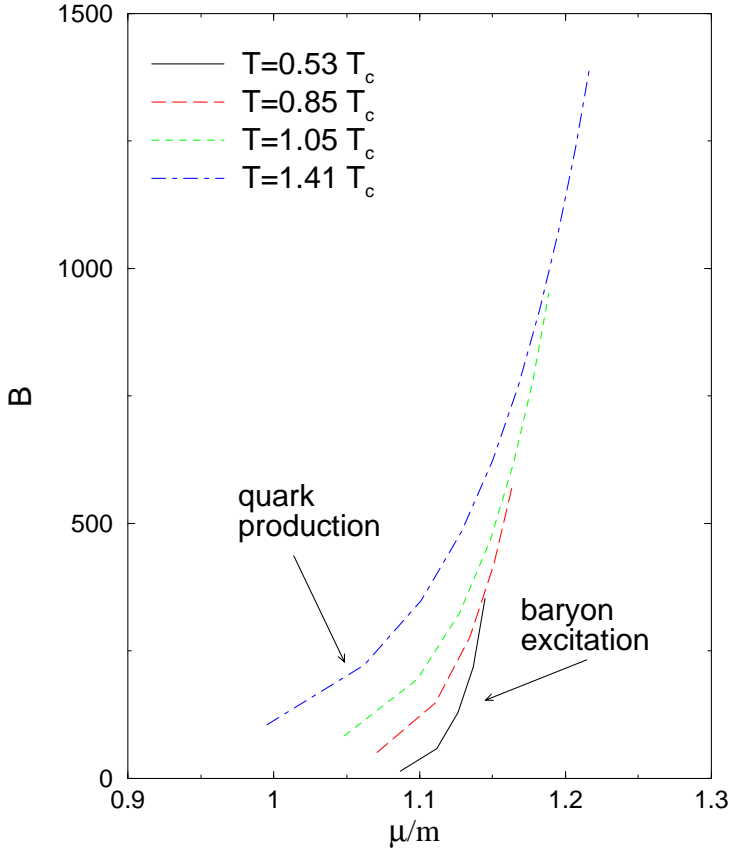} 
}
} \hspace{1cm}
\parbox{4cm}{ 
\caption{ The number $B$ of Baryons which are present in the lattice 
   universe as function of the chemical potential $\mu $ in units 
   of the heavy quark mass $m$. }\label{fig:1} 
}
\end{figure} 
While the derivation of the effective gluonic action in the previous 
subsections holds for a SU($N_c$) gauge group, we will below confine 
ourselves to the case of a SU(2) gauge group. 
The partition function (\ref{eq:3.76}) 
supplemented with the action
\bea 
-S_{latt} &=& \sum _{\{x\}} \biggl( \beta _F \, \sum _{\mu \nu } 
P_{\mu \nu }(x) \; + \; \xi \, \frac{1}{2} \, \tr P(x) \biggr) \; , 
\label{eq:4.1} \\ 
\xi &=& \frac{ 2 m^{3/2} T^{5/2} }{\sqrt{2} \pi ^{3/2} } a^4 \, 
\left[ e^{ \frac{\mu -m }{T} } \; + \; e^{ -  \frac{\mu + m }{T} } 
\, \right] 
\label{eq:4.2} 
\ena 
can be simulated with the standard heat bath algorithm of 
Creutz~\cite{creu80}. The non-locality of the action due to the 
Polyakov loops only yields a modest increase in computational time. 
We adjusted the temperature $T$ of the system by varying the number 
$N_t$ of lattice points in time direction, i.e.,  $T=1/N_t a$. 
In order to remove the superficial renormalization point, i.e.,  
$\beta _f$, dependence from physical quantities, we assume that 
the one-loop scaling (\ref{eq:3.60}) is a good approximation 
for moderate $\beta _f$ values, i.e.,  $\beta _f \in [2.1, 2.5]$. 
In order to fix the overall scales, we used 
\be 
a(\beta _f =2.3 ) \; = \; 0.16 \, \hbox{ fm } \; , 
\label{eq:4.3} 
\en 
which would correspond to a string tension of $\sigma = (440 \, \hbox{MeV} 
)^2 $ in pure Yang-Mills theory $(N_f=0)$. The density of nuclear 
matter is roughly given by $\rho \approx 0.15 \, \hbox{fm}^{-3} $. 
This value corresponds on average to $6 \times 10^{-4}$ Baryons in an 
elementary cube of size $a^3$ and roughly one baryon in a lattice universe 
of size $12 ^3 $. 

\vskip 0.3cm 
In practice, it turned out to be 
convenient to run the simulation for a definite set of $\xi $-values, 
e.g. $\xi \in [0,1/2]$, and to calculate the relation 
between the parameters $\mu $, $m$, $T$ afterwards. 
The derivation of (\ref{eq:4.1}) relies on a truncation of the 
series (\ref{eq:3.73a}) at $n=1$. For being consistent the inequality 
\be 
\frac{1}{ 2 ^{5/2} } \; e^{- m/T } \; \frac{ 
e^{2 \mu / T } \, + \, e^{-2 \mu /T } }{ e^{ \mu / T } \, + \, 
e^{ \mu /T } } \;  \left\langle \vert P \vert \right\rangle \; < \; 1 \; . 
\label{eq:4.4} 
\en 
must be satisfied. In the present paper, we tacitly assume that the 
truncation at $n=1$ is reasonable if (\ref{eq:4.4}) is satisfied. 
Investigations beyond this approximation are left to future work.

\subsection{ Quark genesis versus quark confinement } 

In subsection \ref{sec:3.3}, we have seen that in the deconfined 
phase $(\langle P \rangle = {\cal O}(1) )$ quarks can be produced 
in large numbers if the chemical potential exceeds the value of the 
quark mass. By contrast, in the ''confined'' phase, i.e.,  $\langle P 
\rangle \approx 0$, the production of single quarks by means of the 
chemical potential is forbidden and only thermal excitations of baryons 
(i.e.,  diquarks for the present case of an SU(2) gauge theory) 
are possible. A large number of baryons is expected to occur for 
$\mu > 2m - b$, where $m $ is the quark mass and $b$ is the binding 
energy of the diquark system. In order to study this interplay between quark 
production and confinement we have calculated the number $B$ of baryons 
which are present in the lattice universe as a function of the 
chemical potential for several temperatures. Pure Yang-Mills theory 
$(N_f=0)$ possesses a second order deconfimement phase transition for 
$T = T_c \approx 300 \, $MeV. The result of the simulation 
using a $12^3 \times N_t$ lattice and $\beta _f=2.3$ is shown in 
figure~\ref{fig:1}. For definiteness, we used $m = 10 \sqrt{ \sigma }$. 
We have checked that the inequality (\ref{eq:4.4}) is satisfied for 
the parameter ranges producing figure~\ref{fig:1}. 
For temperatures below $T_c$, we clearly observe 
an onset value $\mu _{onset } $ larger than the quark mass. 
For temperatures larger $T_c$ a significant rise in the baryon number 
is observed for $\mu \approx m$. While in the previous case the 
net baryon number is produced via baryonic excitations, excitations 
of single quarks contribute to the net baryon number in the high 
temperature phase.

\subsection{ String breaking at finite density } 

\begin{figure}[t]
\centerline{ 
\epsfxsize=14cm
\epsffile{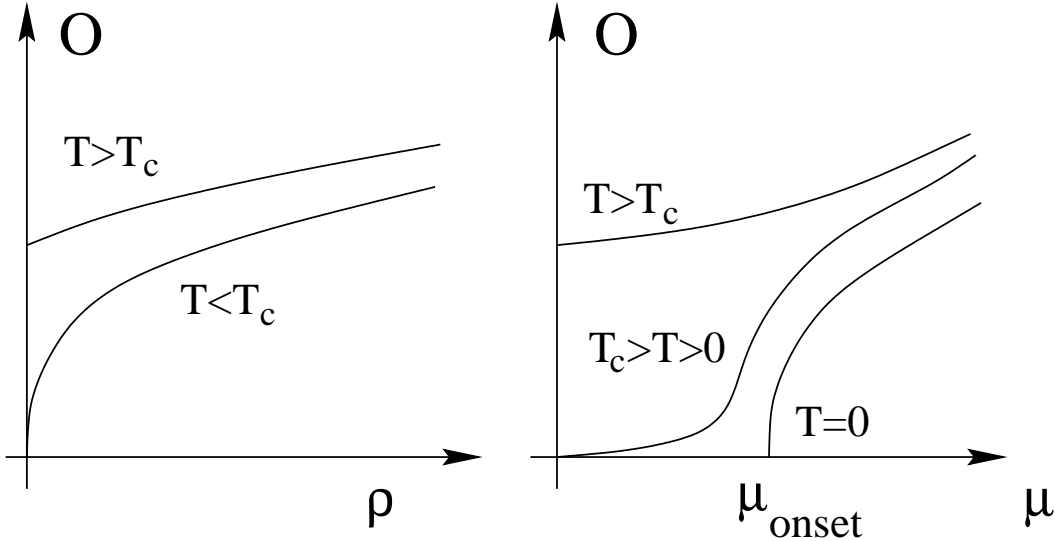} 
}
\caption{ schematic plot: the confinement order parameter 
   (\protect{\ref{eq:4.5}}) of pure SU(2) 
   gauge theory in the case of a finite-density system as a function of 
   the baryon density $\rho $ (left panel) and the chemical potential 
   $\mu $ (right panel), respectively. }\label{fig:2} 
\end{figure} 
At zero baryon (quark) density, a convenient order parameter $O$ 
of confinement is constructed from the Polyakov line $ P(\vec{x})$, 
\be 
O \; := \; \left\langle \left\vert \sum _{\{\vec{x} \}} P(\vec{x}) \right\vert 
\right\rangle \; . 
\label{eq:4.5} 
\en 
Keeping in mind that the Polyakov line reverses its sign under a center 
transformation, i.e.,  $P(\vec{x}) \rightarrow - P(\vec{x}) $, 
$\langle P(\vec{x}) \rangle =0 $ is mandatory for a realization of 
center symmetry. In this case, quark confinement is 
inherent~\cite{sve86} and one 
observes $ O \propto 1/ \sqrt{V}$, where $V$ is the space volume. 
If the temperature exceeds the critical temperature $T_c$ ($\approx 
300 \, $MeV for a pure SU(2) gauge theory), a non-vanishing 
expectation value of the Polyakov line, i.e.,  $\langle P(x) \rangle 
\not=0 $ and therefore $O = {\cal O}(1)$, signals the spontaneous 
breakdown of center symmetry and deconfinement. 

\vskip 0.3cm 
While at zero density 
the color electric string between static quark sources might extend 
to arbitrary length, one expects at finite densities a maximum length $l_0$ 
of the color electric string which is controlled by the average distance 
between the background quarks. In this case, string breaking occurs and 
the heavy quark potential saturates at a quark distance $l_0$. A 
precise definition of a deconfinement order parameter is cumbersome 
in the finite density case, and has been under debate for more than twenty 
years (for recent progress see~\cite{satz98}). 

\vskip 0.3cm 
Although one does not expect a sharp deconfinement phase 
transition as a function of temperature at finite densities, 
the gluonic state might undergo drastic changes which are triggered 
by temperature effects and which are signaled by significant (but smooth) 
changes in observables. To these respects, the behavior 
of $O$ is of particular interest also at finite densities. 
In the present case of an SU(2) gauge group
($\tr P = \tr P^\dagger $) and a system of heavy quarks, the density 
(\ref{eq:3.78}) is proportional to the expectation value of the 
Polyakov loop. One therefore finds 
\be 
\vert \rho \vert \; \propto \; \left\vert \left\langle \tr P 
\right\rangle \right\vert \; \le \; \left\langle \left\vert \tr P 
\right\vert \right\rangle \; , 
\label{eq:4.6} 
\en 
and concludes that $O = {\cal O}(1)$ as long as $\rho \not=0 $. 
Our result nicely confirms the observation in~\cite{kar99} that color 
electric string breaking occurs as soon as the baryon density is 
non-zero. 

\vskip 0.3cm 
We are led to the following behavior of $O$ as function of the 
chemical potential and density, respectively: for temperatures 
significantly below the critical temperature and for $\mu < \mu _{onset}$, 
the baryon density is practically zero yielding small values of 
$O$. If the chemical potential $\mu $ exceeds $\mu _{onset}$, 
the drastic rise of the density is accompanied by a strong increase 
of $O$. At large temperatures $(T>T_c)$, $O$ is non-zero due to 
temperatures effects and changes in $O$ due to density effects are 
moderate in this case (see figure~\ref{fig:2}). 
Numerous numerical results confirm the qualitative behavior shown in 
figure~\ref{fig:2}. Quantitative details are not interesting since they 
strongly depend on the actual choice $m \gg \sqrt{\sigma }$ and on the 
fine tuning $\mu \rightarrow m ^+$.

\section{Conclusions} 

Our central idea of the present paper is to combine analytic 
methods for calculating the fermion determinant with the lattice 
description for deriving a valuable description of the Yang-Mills system 
at finite densities. Removing the ultra-violet regularization 
($\Lambda \rightarrow \infty $ in the case of the quark determinant 
and $ a \rightarrow 0$ in the case of the gluonic functional integral), 
physical quantities are independent of the type of regularization 
and approach a unique result. 
We have shown in section 3.2 how the proper scaling towards the 
continuum limit is obtained in the present case of interest. 
The advantage of our approach is twofold: firstly, by construction the 
approach is not plagued with spurious quark states (see e.g.~\cite{ste96}). 
Secondly, the physical heavy mass limit, i.e., $\Lambda _{YM}, \, T \ll m \ll 
\Lambda $, is manifest. 

\vskip 0.3cm 
In the case of the grand-canonical partition function, the chemical 
potential is chosen of the order of the quark mass in order to 
produce significant effects in the quark density~\cite{blu96}. 
Our first numerical results for the case of an SU(2) gauge group have been 
presented in section 4. Rising the temperature above the deconfinement 
critical one we observe a decrease of the onset value of the chemical 
potential at which a rapid increase of baryon density is observed. 
We interpret this result as follows: at low temperatures, baryon 
density is generated by the chemical potential only via the production of 
baryons (i.e., diquarks in the case of an SU(2) gauge group). At high 
temperatures, by contrast, the excitation of single quarks contributing to 
the density becomes feasible due to deconfinement. 

\vskip 0.3cm 
In the case of the canonical partition function $Z(B)$, describing 
a system with baryon number $B$, we observe that the quark 
determinant is expressed in terms of products of Polyakov loops. 
In agreement with the findings in~\cite{kar99}, the determinant 
is center symmetric, and a non-vanishing expectation value of the Polyakov 
loop at finite densities occurs via string breaking~\cite{kar99}. 

\vskip 0.3cm 
Although the calculation of the quark determinant in the present paper 
is tied to the Schwinger proper-time approach which generates the 
large mass expansion, the basic idea of combining an analytic calculation 
of the determinant with a subsequent lattice representation of the 
gluon fields is quite universal. Estimating fermion determinants 
by resorting to different types of approximation schemes 
has a long history in the literature. The idea of studying the 
opposite limit $m \rightarrow 0 $ (chiral limit) by applying this 
new idea seems very appealing to us.

\vskip 1cm 
{\bf Acknowledgments: } 

We thank Mannque Rho and Dong-Pil Min for helpful discussions and 
encouragement. KL is indebted to Holger Gies for interesting discussions 
on fermion determinants in Schwinger proper-time regularization. 
We greatly acknowledge the hospitality of KIAS where large parts of the 
present work was performed. We thank C.~W.~Kim, President 
of KIAS, who made this collaboration possible.

\appendix
\setcounter{equation}{0}\renewcommand{\theequation}{\mbox{A.\arabic{equation}}}
\section{ Notation and conventions }
\label{app:a}

The metric tensor in Minkowski space is
\be
g_{\mu \nu } \; = \; \hbox{diag} ( 1 , -1 , -1 , -1 ) \; .
\label{eq:aa1}
\en
We define Euclidean tensors $T_{(E)}$ from the tensors in Minkowski
space $T_{(M)}$ by
\be
T^{\mu _1 \ldots \mu _N }_{ (E) \phantom{\ldots \mu _N } \nu _1
\ldots \nu _n }
\; = \; (i)^r \, (-i)^s \;
T^{\mu _1 \ldots \mu _N }_{ (M) \phantom{\ldots \mu _N } \nu _1
\ldots \nu _n } \; ,
\label{eq:aa2}
\en
where $r$ and $s$ are the numbers of zeros within $\{ \mu _1 \ldots
\mu _N \}$ and $\{ \nu _1 \ldots \nu _n \} $, respectively.
In particular, we have for the Euclidean time and the Euclidean
metric
\be
x_{(E)} ^ 0 \; = \; i \, x_{(M)}^0 \; , \hbo
g^{\mu \nu }_{(E)} \; = \; \hbox{diag} ( -1 ,-1 ,-1 ,-1 ) \; .
\label{eq:aa3}
\en
Covariant and contra-variant vectors in Euclidean space differ by an
overall sign. For a consistent treatment of the symmetries, one is forced
to consider the $\gamma ^\mu $ matrices as vectors. Therefore, one is
naturally led to anti-hermitian Euclidean matrices via (\ref{eq:aa2}),
\be
\gamma ^0 _{(E)} = i \gamma ^0 _{(M)} \; , \hbo
\gamma ^k _{(E)} = \gamma ^k _{(M)} \; .
\label{eq:aa4}
\en
In particular, one finds
\be
\left( \gamma ^\mu _{(E)} \right) ^{\dagger } \; = \;
- \; \gamma ^\mu _{(E)} \; , \hbo
\{ \gamma ^\mu _{(E)} , \gamma ^\nu _{(E)} \} \; = \;
2 g^{\mu \nu }_{(E)} \; = \; -2 \, \delta _{\mu \nu } \; .
\label{eq:aa5}
\en
The so-called Wick rotation is performed by considering the
Euclidean tensors (\ref{eq:aa2}) as real fields.

\vskip 0.3cm
In addition, we define the square of an Euclidean
vector field, e.g. $V_\mu $, by
\be
V^2 \; := \; V_\mu V_\mu \; = \; - V_\mu V^\mu \; .
\label{eq:aa51}
\en
This implies that $V^2$ is always a positive quantity (after the wick
rotation to Euclidean space).

\setcounter{equation}{0}\renewcommand{\theequation}{\mbox{B.\arabic{equation}}}
\section{ The off-diagonal heat coefficients } 
\label{app:b}

\begin{figure}[t]
\centerline{ 
\epsfxsize=12cm
\epsffile{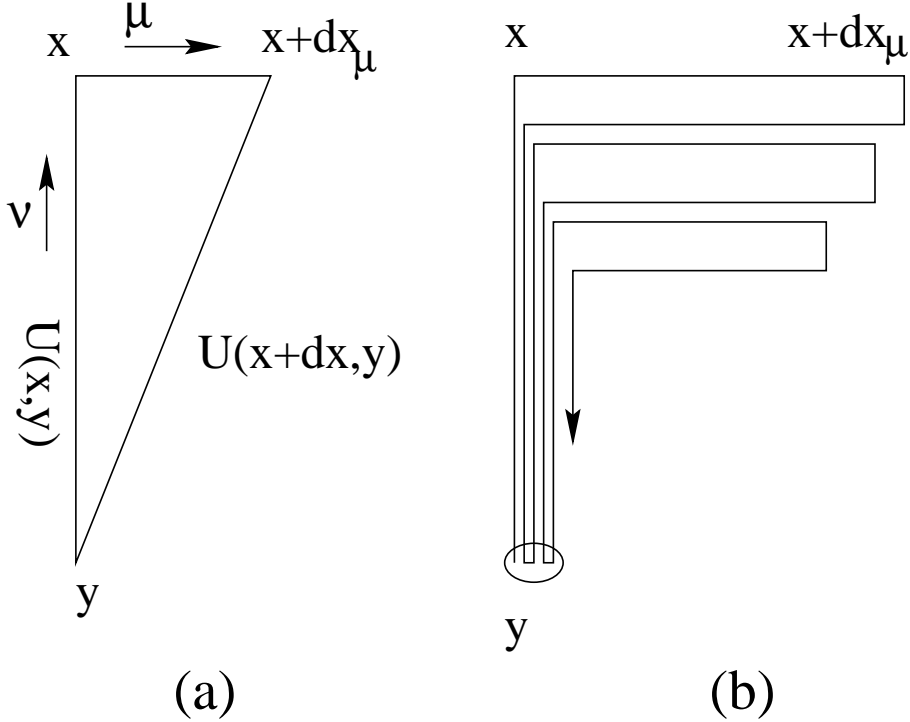} 
}
\caption{ Possibilities (a) and (b) for calculating the holonomy 
  along the triangle path shown in (a) (non-Abelian Stokes theorem). 
  }\label{fig:3} 
\end{figure} 
The aim of this subsection is to calculate the full space time dependence 
of the heat coefficients $h_k(x,y)$. We will show that we recover 
the well known result for the diagonal part $h_k(x,x)$. For this purpose, 
our starting point is the recursion relation 
\be
(k+z_\mu D_\mu)h_k(x,y)=(D^2+{i\over2}\sigma_{\mu\nu}[D_\mu,D_\nu]) 
h_{k-1}(x,y) \; , 
\label{eq:b1}
\en
while the equation for $h_0(x,y)$ is given by 
\be
z_\mu D_\mu h_0(x,y)=0 \; , \hbo z_\mu \, := \, x_\mu - y_\mu \; . 
\label{eq:b2}
\en

We first show that the gauge covariant connection 
\be 
h_0(x,y) \; = \; U(x,y) \; = \; {\cal P} \exp \left\{ -i \int_{C_{xy}} 
A_\mu (x^\prime ) \; dx_\mu ^\prime \, \right\} 
\label{eq:b3}
\en
where the path $C_{xy}$ is a straight line connecting the points 
$x$ and $y$, provides a solution to equation (\ref{eq:b2}). 
For a proof we calculate the holonomy along the triangle path 
shown in figure~\ref{fig:3}. 
By a comparison of the results obtained by using the equivalent 
paths (a) and (b) in figure~\ref{fig:3} at the level ${\cal O } (dx)$, 
one finds 
\bea 
U^\dagger (x,y) \, D_\mu U(x,y) &=& -i \, \int_0^1 \; dt \; t \, z_\nu 
\; {\cal F} _{\mu \nu } (zt+y,y ) \; , 
\label{eq:b4} \\ 
{\cal F} _{\mu \nu } (x^\prime , y ) &:=& U^\dagger (x^\prime , y) \; 
F _{\mu \nu } (x^\prime ) \; U (x^\prime , y) \; . 
\label{eq:bb4}
\ena    
In the Abelian case ${\cal F} _{\mu \nu } (x^\prime , y )$ becomes 
independent of $y$ and coincides with the standard field strength 
$F _{\mu \nu } (x^\prime )$. It is can be easily checked that both sides 
of (\ref{eq:b4}) transform homogeneously under gauge transformations. 
Using (\ref{eq:bb4}) and the anti-symmetry of the tensor 
${\cal F} _{\mu \nu } (x^\prime , y )$ under an exchange of the indices 
$\mu $ and $\nu $, one immediately observes that 
$ z_\mu D_\mu h_0(x,y) = 0$. 

\vskip 0.3cm 
In order to calculate the heat coefficients $h_k(x,y)$, we decompose 
$$ 
h_k(x,y) \; = \; U(x,y) \, g_k(x,y)  \; , \hbox to 3cm {\hfill with 
\hfill } g_0(x,y) \; = \; 1 \; . 
$$
Making extensive use of (\ref{eq:b4}) and the relation 
$$ 
D_\mu \, \biggl( U(x,y) \, g_k(x,y) \biggr) \; = \; 
\biggl( D_\mu \, U(x,y) \biggr) \, g_k(x,y) \; + \; U(x,y) \, 
\biggl(  \partial _\mu g_k(x,y) \biggr) \; , 
$$ 
one can cast (\ref{eq:b1}) into a recursion relation for $g_k(x,y)$, i.e.,  
\bea
(k+z_\mu \partial_\mu)\lefteqn{g_k(x,y)}\nonumber \\
    &=&\left(\partial^2-{2i\over z}\int_{C_{xy}}
    z'dx'_\alpha {\cal F} _{\mu\alpha}(x',y)\partial_\mu\right. \nonumber\\
    &&-{1\over z^2}\int_{C_{xy}}z'dx'_\alpha {\cal F} _{\mu\alpha}(x',y)
    \int_{C_{xy}}z''dx''_\beta {\cal F} _{\mu\beta}(x'',y)\nonumber\\
    &&-{i\over z^2}\int_{C_{xy}}z'^2dx'_{\alpha}
     \partial'_\mu {\cal F} _{\mu\alpha}(x',y)
    -{\sigma_{\mu\nu}\over z^2}\int_{C_{xy}}z'_\alpha dx'_\alpha 
    {\cal F} _{\mu\nu}(x',y)      \nonumber \\
    &&-{i\sigma_{\mu\nu}\over 2z^2}\int_{C_{xy}}z' dx'_\alpha  
    \int_{C_{xy}}z'' dx''_\beta
    [{\cal F} _{\mu\alpha}(x',y),{\cal F} _{\nu\beta}(x'',y)]\nonumber\\
    &&+\left. {\sigma_{\mu\nu}\over 2z^2}\int_{C_{xy}}z'^2dx'_\alpha 
    (\partial'_\mu {\cal F} _{\nu\alpha}(x',y)-\partial'_\nu 
    {\cal F} _{\mu\alpha}(x',y))\right)g_{k-1}(x,y) \; . 
\label{eq:b5}
\ena      
In particular, the equation for $g_1(x,y)$ becomes 
\bea
(1+z_\mu\partial_\mu)g_1(x,y
   \lefteqn{)=-{1\over z^2}\int_{C_{xy}}z'dx'_\alpha 
    {\cal F} _{\mu\alpha}(x',y)
    \int_{C_{xy}}z''dx''_\beta {\cal F} _{\mu\beta}(x'',y)}\nonumber\\
      &&-{i\over z^2}\int_{C_{xy}}z'^2dx'_{\alpha}
     \partial'_\mu {\cal F} _{\mu\alpha}(x',y) 
     -{\sigma_{\mu\nu}\over z^2}\int_{C_{xy}}z'_\alpha dx'_\alpha 
     {\cal F} _{\mu\nu}(x',y)
        \nonumber \\
     &&-{i\sigma_{\mu\nu}\over 2z^2}\int_{C_{xy}}z' dx'_\alpha 
       \int_{C_{xy}}z'' dx''_\beta
       [{\cal F} _{\mu\alpha}(x',y),{\cal F} _{\nu\beta}(x'',y)] 
\nonumber\\
     &&+ {\sigma_{\mu\nu}\over 2z^2}\int_{C_{xy}}z'^2dx'_\alpha (\partial'_\mu
      {\cal F} _{\nu\alpha}(x',y)-\partial'_\nu {\cal F} _{\mu\alpha}(x',y))
\label{eq:b6}
\ena
Referring to the particular solution of the equation 
$(k+z_\mu\partial_\mu)A(x,y)=B(x,y)$ given by 
\be
A(x,y)={1\over z^k}\int_{C_{xy}}z'^{k-2}\,z'^\mu B(x,y)dx'^\mu 
\label{eq:b7}
\en
we finally obtain
\bea
g_1(x,y)=-{1\over z}\int_{C_{xy}}z'^{-1}z'_\lambda dx'_\lambda
        \lefteqn{\left({1\over z'^2}\int_{C_{x'y}}z''dx''_\alpha 
        {\cal F} _{\mu\alpha}(x'',y)
       \int_{C_{x'y}}z'''dx'''_\beta {\cal F} _{\mu\beta}(x''',y )\right. } 
\nonumber\\
     &&\hspace{-4mm}  +{i\over z'^2}\int_{C_{x'y}}z''^2 dx''_{\alpha}
       \partial''_\mu {\cal F} _{\mu\alpha}(x'',y)\nonumber\\
     &&\hspace{-4mm}+{\sigma_{\mu\nu}\over z'^2}
     \int_{C_{x'y}}z''_\alpha dx''_\alpha {\cal F} _{\mu\nu}(x'',y)
\label{eq:b8} \\ 
    &&\hspace{-4mm}+{i\sigma_{\mu\nu}\over 2z'^2}
     \int_{C_{x'y}}z'' dx''_\alpha\int_{C_{x'y}}z''' dx'''_\beta
          [{\cal F} _{\mu\alpha}(x'',y),{\cal F} _{\nu\beta}(x''',y)]
\nonumber\\
    &&\hspace{-4mm}-\left. {\sigma_{\mu\nu}\over 2z'^2}
    \int_{C_{x'y}}z''^2dx''_\alpha (\partial''_\mu
     {\cal F} _{\nu\alpha}(x'',y)-\partial''_\nu 
     {\cal F} _{\mu\alpha}(x'',y))\right)
\nonumber 
\ena 
Not showing terms which vanish if the trace over Dirac indices 
is performed, we find 
\bea
h_1(x,y)& =& -{\cal P } \, \exp \left\{ -i \int _{C_{xy}}
A_\mu (x^\prime ) \; dx_\mu ^\prime \, \right\}
\frac{1}{z}\int_{C_{xy}}z'^{-1}z'_\lambda \; dx'_\lambda
\nonumber \\
&& \frac{1}{z'^2}\int_{C_{x'y}}z''\; dx''_\alpha\;
{\cal F} _{\mu\alpha}(x'',y)\int_{C_{x'y}}z'''\; dx'''_\beta\;
{\cal F} _{\mu\beta}(x''',y) \; + \; \ldots \; . 
\label{eq:h1}
\ena

For calculating $g_2(x,y)$, we introduce 
$$
x'=zp+y,\,x''=z't+y=zpt+y,\,x'''=z's+y=zps+y
$$
and rewrite $g_1(x,y)$ as 
\bea
g_1(x,y)=\lefteqn{-\int_0^1p^2 dp\int_0^1 tdt\int_0^1sds z_{\alpha}
            z_\beta {\cal F} _{\mu\alpha}(zpt+y,y) 
            {\cal F} _{\mu\beta}(zps+y,y )}
\nonumber \\
        &&-i\int_0^1dp\int_0^1tdt z_\alpha\partial_\mu 
           {\cal F} _{\mu\alpha}(zpt+y,y )
\nonumber\\
        &&-\sigma_{\mu\nu}\int_0^1dp\int_0^1tdt 
           {\cal F} _{\mu\nu}(zpt+y,y)
\nonumber\\
        &&-{i\sigma_{\mu\nu}\over2}\int_0^1p^2 dp\int_0^1tdt\int_0^1sds
         z_\alpha z_\beta[{\cal F} _{\mu\alpha}(zpt+y,y), 
         {\cal F} _{\nu\beta}(zps+y,y)]
\nonumber\\
        &&+{\sigma_{\mu\nu}\over2}\int_0^1dp\int_0^1tdt z_\alpha
        (\partial_\mu {\cal F} _{\nu\alpha}(zpt+y,y)-\partial_\nu 
        {\cal F} _{\mu\alpha}(zpt+y,y))
\label{eq:b9}
\ena
where $\partial_\mu {\cal F} _{\mu\nu}(zpt+y,y)=\partial_\mu 
{\cal F} _{\mu\nu}(x,y)|_{x=zpt+y} $. 
Using (\ref{eq:b5}), (\ref{eq:b7}) and (\ref{eq:b9}), 
the final result of a lengthy calculation is 
\bea
g_2(x,y)\lefteqn{=\frac{1}{z^2}\int_{C_{xy}}z'_\gamma\;
         dx'_\gamma \left\{-\int_0^1p^2 dp\int_0^1 tdt\int_0^1sds
          \biggl(2{\cal F} _{\mu\nu}(z'pt+y,y) {\cal F} _{\mu\nu}(z'ps+y,y)
         \right. }
\nonumber\\
           &&\hspace{3cm}+2z'_\beta\partial'_\lambda
              (\{{\cal F} _{\mu\lambda}(z'pt+y,y), 
              {\cal F} _{\mu\beta}(z'ps+y,y)\})
\nonumber\\
              &&\hspace{3cm}+z'_{\alpha}z'_\beta\partial'^2
             {\cal F} _{\mu\alpha}(z'pt+y,y) {\cal F} _{\mu\beta}(z'ps+y,y) 
             \biggr)
\nonumber \\
         &&\hspace{-4mm}+2\left(\int_0^1dp\int_0^1tdt 
            {\cal F} _{\mu\nu}(z'pt+y,y) \right. 
\nonumber\\
        &&+{i\over2}\int_0^1p^2 dp\int_0^1tdt\int_0^1sds
         z'_\alpha z'_\beta[{\cal F} _{\mu\alpha}(z'pt+y,y), 
         {\cal F} _{\nu\beta}(z'ps+y,y)] 
\nonumber\\
        &&-\left.{1\over2}\int_0^1dp\int_0^1tdt z'_\alpha
      (\partial'_\mu {\cal F} _{\nu\alpha}(z'pt+y,y)-\partial_\nu 
      {\cal F} _{\mu\alpha}(z'pt+y,y))
       \right)^2\biggr\} \; + \; \ldots \; , 
\label{eq:b10}
\ena
where we have not shown the traceless terms. 
In the diagonal limit $ x \rightarrow y$, the formulas for 
$g_1(x,y)$ and $g_2(x,y)$ greatly simplify and we recover the 
familiar result 
\bea
g_1(x,x)&=&0\nonumber\\ 
g_2(x,x)&=&-{1\over12}F^2+{1\over8}F^2\nonumber\\
   &=&{1\over6}F^2 \; . 
\label{eq:b11}
\ena

\end{document}